\documentclass[article,showpacs,amsmath,amssymb,nofootinbib]{revtex4}
\usepackage{graphicx}

\usepackage{dcolumn}
\usepackage{bm}
\makeatother

\begin{document}

\title{Bosonization and Quantum Hydrodynamics}
\author{Girish S. Setlur}
\affiliation{Department of Physics, Indian Institute of
Technology, Guwahati, \\North Guwahati, Assam 781 039, India}

\date{\today}

\begin{abstract}

It is shown that it is possible to bosonize fermions in any number
of dimensions using the hydrodynamic variables, namely the
velocity potential and density.  The slow part of the Fermi field
is defined irrespective of dimensionality and the commutators of
this field with currents and densities are exponentiated using the
velocity potential as conjugate to the density.  An action in
terms of these canonical bosonic variables is proposed that
reproduces the correct current and density correlations. This
formalism in one dimension is shown to be equivalent to the
Tomonaga-Luttinger approach as it leads to the same propagator and
exponents. We compute the one-particle properties of a spinless
homogeneous Fermi system in two spatial dimensions with long-range
gauge interactions and highlight the metal-insulator
 transition in the system.
 A general formula for the generating function of density
 correlations is derived that is valid beyond the random phase
 approximation. Finally, we write down a formula for
 the annihilation operator in momentum space directly in terms of
 number conserving products of Fermi fields.
\ \\
\end{abstract}



 \maketitle

\section{Introduction}

The study of interacting fermions in one spatial dimension has
attracted renewed attention in the recent past with the
experimental realization of nano-structures and simulation of
quasi-one dimensional environments\cite{Expt1}. Many theoretical
works that seek to provide a basis for testing fundamental
physics, as well as those that propose new scenarios for practical
application of these ideas have emerged\cite{Expt2}. All these are
based on the Luttinger liquid paradigm of Haldane\cite{Haldane}
where homogeneous one dimensional systems are considered to be a
specific kind of anomalous metal characterized by power law
singularities in the Green function. However, this paradigm is
valid mainly for short-range interactions. For a specific kind of
long-range interaction ($ v(q) = 2 e^2/q^2$), one of the authors
have shown in an earlier work\cite{PraLutt} that the system is
characterized by essential singularities rather than power-law
singularities. Also, in another work\cite{Setlur1}, the formal
theoretical framework upon which the Tomonaga-Luttinger picture is
based, was critiqued. In particular, this approach fails to
provide a proper description of the ground state of a one
dimensional homogeneous Fermi system interacting via the gauge
potential as has been pointed out \cite{PraLutt}.  That the
presence of Klein factors in this approach, renders the analysis
of the three-wire junction problematic has been noted by Chamon et
al \cite{Chamon} and by Lal et al\cite{Expt2}.

In this paper we show that fermions in one and higher dimensions
may be conveniently bosonized by exponentiating the slow part of
the field using current algebra.  In particular, it is shown that
no Klein factors need be invoked \footnote{More precisely, just
one global Klein factor suffices in all dimensions rather than one
for each kind of mover\cite{Setlur1}, \cite{Citrin}.}.
 The right movers
and left movers are shown to obey fermion commutation rules
nonetheless.
 In one dimension, this fact is already known to the experts\cite{Shankar}\cite{Delft}, however
 in higher dimensions it is probably less well known. Nevertheless, we present the one dimensional
 formalism even if there is some overlap with established methods, if only to motivate
 the generalization to higher dimensions.
 We also express the annihilation operator in momentum
space directly in terms of a certain combination of number
conserving products of Fermi fields known as
sea-bosons\cite{Setlur1} thereby completing the formalism
introduced earlier\cite{Citrin}.
 It is worth emphasizing that it is the {\it{slow}} part of the
  field operator that is easily expressed in terms of the bosons
  rather than the full field as was implied in one of the authors'
  earlier work\cite{Setlur1}.
 The present approach, together with the
action mentioned in the abstract, can be argued to be quite
adequate in studying gauge interactions in higher dimensions as
well.

\section{Formalism}

For a pedagogical introduction to bosonization in one dimension,
there are many reviews available\cite{Delft}.
For the sake of brevity, we shall only discuss the important
ideas. First we consider fermions in one spatial dimension. Define
the right movers and left movers as usual ($ \Lambda \ll k_{F} $
is the momentum cutoff around the Fermi points), $ \psi_{R}(x) =
\frac{1}{\sqrt{L}} \sum_{ k } e^{ikx} c_{ k_{F}+k }
\theta(\Lambda-|k|) $ and $ \psi_{L}(x) = \frac{1}{\sqrt{L}}
\sum_{ k } e^{ikx} c_{ -k_{F}+k } \theta(\Lambda-|k|)$. These are
slowly varying in space. In this theory we assume that in the end
$ k_{F} \gg \Lambda \rightarrow \infty $.
 Now we define the density fluctuation operator and
its conjugate as follows ( with $ p $ and $ q $ unrestricted ), $
\rho_{q} = \sum_{ p } c^{\dagger}_{p + q/2}c_{p-q/2}$ and $ X_{q}
= \sum_{ p } \left( \frac{ i p.q }{ q^2 N^{0} } \right) \mbox{
}c^{\dagger}_{p-q/2}c_{p+q/2} $. These operators obey the
following commutation relations ,$ [X_{q}, X_{q^{'}}] = [\rho_{q},
\rho_{q^{'}}] = 0 $ and $ [X_{q}, \rho_{q^{'}}] = i\mbox{
}\delta_{q,q^{'}} $\cite{Setlur1}. Since only terms  with small $
q $ contribute ( the $ \psi $ 's are slowly varying ) we have,$
[\psi_{R}(x),\rho_{q}] = e^{iqx} \psi_{R}(x)$,$ [\psi_{R}(x),
X_{q}] =  e^{-iqx}  \left( \frac{ i \pi }{ q L } \right)
\psi_{R}(x)$ and $ [\psi_{L}(x),\rho_{q}] = e^{iqx} \psi_{L}(x) $,
$ [\psi_{L}(x), X_{q}] = -\mbox{  } e^{-iqx} \left( \frac{ i \pi
}{ q L } \right) \psi_{L}(x)$. These may be exponentiated as
follows.
\begin{equation}
\psi_{R}(x) = e^{ -i \sum_{q} e^{iqx} X_{q} }
 e^{ -\pi \sum_{q} \frac{ e^{-iqx} }{qL} \rho_{q} } \mbox{           }\sqrt{ \frac{
 \Lambda }{ 2 \pi } }
\label{EQPSIR}
\end{equation}
\begin{equation}
\psi_{L}(x) = e^{ -i \sum_{q} e^{iqx} X_{q} }
 e^{ \pi \sum_{q} \frac{ e^{-iqx} }{qL} \rho_{q} } \mbox{           }\sqrt{ \frac{
 \Lambda }{ 2 \pi } }
\label{EQPSIL}
\end{equation}
The multiplicative factor, namely $\sqrt{\frac{\Lambda}{2\pi}}$,
comes from the observation that, $ \left< \psi^{\dagger}_{R}(x)
\psi_{R}(x) \right>
 = \left< \psi^{\dagger}_{L}(x) \psi_{L}(x) \right> = \frac{\Lambda
 }{2\pi} $.
The Eq.(\ref{EQPSIR}) and Eq.(\ref{EQPSIL}) obey fermion
commutation rules without the need for Klein factors for $ x \neq
x^{'} $ which is easy to verify. In the usual approach to
bosonization, these formulae are regularized by introducing a
decaying exponential\cite{Delft}. The regularization is needed to
reproduce the delta function in the fermion commutation rules,
  $ \{ \psi_{R/L}(x), \psi^{\dagger}_{R/L}(x^{'}) \} = \delta(x-x^{'}) $.
Here too, we adopt a similar but not quite the same approach. In
this regard we postulate,
\begin{equation}
\psi_{R}(x) = e^{ -i \sum_{q} e^{iqx} X_{q} }
 e^{ -\pi \sum_{q} \frac{ e^{-iqx} }{qL} \rho_{q} }
 \mbox{           } \mbox{           }\left( \frac{
 \Lambda }{ 2 \pi }  + \frac{1}{L} \sum_{q \neq 0 } e^{ -i q x }
 [\frac{1}{2}\rho_{q} + \frac{ i q L }{ 2\pi } X_{-q}] \right)^{\frac{1}{2}}
\end{equation}
\begin{equation}
\psi_{L}(x) =  e^{ -i \sum_{q} e^{iqx} X_{q} }
 e^{ \pi \sum_{q} \frac{ e^{-iqx} }{qL} \rho_{q} }
 \mbox{           }\left( \frac{
 \Lambda }{ 2 \pi }  + \frac{1}{L} \sum_{q \neq 0 } e^{ -i q x }
 [\frac{1}{2}\rho_{q} - \frac{ i q L }{ 2\pi }
X_{-q}] \right)^{\frac{1}{2}}
\end{equation}
That is, we add the density fluctuations of right/left movers to
the (nearly infinite) mean density. This means that the
commutation relations with $ \rho_{q} $ and $ X_{q} $ are now
going to be invalid but the corrections are inverse in the number
of right and left movers. Since $ \Lambda \rightarrow \infty $ we
may write,
\begin{equation}
\psi_{R}(x) = e^{ -i \sum_{q \neq 0} e^{iqx} X_{q} }
 e^{ -\pi \sum_{q \neq 0} \frac{ e^{-iqx} }{(q-i\epsilon)L} \rho_{q} }
 \mbox{           } \mbox{           }e^{ \frac{\pi}{L\Lambda}
  \sum_{q \neq 0 } e^{ -i q x }e^{ -|q|/\Lambda }
 [\frac{1}{2}\rho_{q} + \frac{ i q L }{ 2\pi } X_{-q}] }
\left( \frac{ \Lambda }{ 2 \pi } \right)^{\frac{1}{2}}
\label{PSIR}
\end{equation}
\begin{equation}
\psi_{L}(x) =  e^{ -i \sum_{q \neq 0 } e^{iqx} X_{q} }
 e^{ \pi \sum_{q \neq 0} \frac{ e^{-iqx} }{(q-i\epsilon)L} \rho_{q} }
 \mbox{           }e^{ \frac{ \pi}{ \Lambda L} \sum_{q \neq 0 }
  e^{ -i q x }e^{ -|q|/\Lambda }
 [\frac{1}{2}\rho_{q} - \frac{ i q L }{ 2\pi }
X_{-q}] } \left( \frac{ \Lambda }{ 2 \pi } \right)^{\frac{1}{2}}
\label{PSIL}
\end{equation}
where $ \epsilon \rightarrow 0^{+} $. It can be shown that
Eq.(\ref{PSIR}) and Eq.(\ref{PSIL}) obey the following commutation
relations,
\begin{equation}
\{\psi_{R}(x), \psi_{L}(x^{'})\} = \{\psi_{R}(x),
\psi^{\dagger}_{L}(x^{'})\} = \{\psi_{R}(x), \psi_{R}(x^{'})\}
 =  \{\psi_{L}(x), \psi_{L}(x^{'})\} = 0
\end{equation}
\begin{equation}
\{\psi_{R}(x), \psi^{\dagger}_{R}(x^{'})\}  = \{\psi_{L}(x),
\psi^{\dagger}_{L}(x^{'})\} = \delta(x-x^{'})
\end{equation}
 These nuances are not very important for the practical
 computations. In fact, we may use the unregularized fields
 for computing the propagators with finite spatial and temporal
 separations with impunity. We remarked that there are no Klein
 factors in the formalism. This refers to the need to invoke two different Klein
 factors, one for each kind of mover,
  in the usual approach\cite{Delft}. In the present approach, there has to
 be an overall global Klein factor namely $ X_{q=0} $ that was
 defined in our earlier work\cite{Citrin}.
We may now compute the propagators as follows. It can be shown
that for the free Fermi theory, $ \left<
\psi^{\dagger}_{R}(x^{'},t^{'}) \psi_{R}(x,t) \right>
 \sim \frac{1}{ (x-x^{'}) - v_{F}(t-t^{'}) } $.
One may also similarly study the interacting system with equal
ease and obtain
 the right exponents for the Luttinger model.

Next, we wish to generalize these ideas to higher dimensions. This
is likely to be complicated since there is nothing called a right
mover or left mover in two and more dimensions because the Fermi
surface is no longer made of two points but rather, an uncountable
infinity of them. The usual approach of Luther and
Haldane\cite{Haldane} breaks up the Fermi surface into patches
where the separation between the patches $ \Delta k_{F} \gg
\Lambda \sim \Delta q $.  Not only is this contrived, it involves
the introduction of one Klein factor in every radial direction.
Here we show that there is no need for Klein factors at all (apart
from the global one\cite{Citrin}). To this end we have to make the
following analogy. Consider the following operators in one
dimension, $ \psi_{slow}(x) = \frac{1}{\sqrt{L} } \sum_{k} e^{ i
(k-k_{F}sgn(k))x } \mbox{ }c_{ k } \mbox{ }\theta(\Lambda - | |k|
- k_{F} | ) $
 and $ \psi_{diff}(x) = \frac{1}{ \sqrt{L} } \sum_{k} e^{ i
(k-k_{F}sgn(k))x } \mbox{       }c_{ k } sgn(k)\mbox{
}\theta(\Lambda - | |k| - k_{F} | ) $. These are linear
combinations of right and left movers which can be obtained from,
\begin{equation}
\psi(x,y) = \frac{1}{ \sqrt{L} } \sum_{k} e^{ i (k-k_{F}sgn(k))x }
e^{ i sgn(k) y }\mbox{       }c_{ k } \mbox{
 }\theta(\Lambda - | |k| - k_{F} | )
\end{equation}
as $ \psi_{slow}(x) = \psi(x,0) \mbox{           }; \mbox{ }
\psi_{diff}(x) =  -i \psi_{y}(x,0) $. By analogy we may generalize
these ideas to more than one dimension as follows. We make the
following prescription:
\begin{math} sgn(k) \rightarrow {\hat{k}}
\end{math}.
Then,
\begin{equation}
\psi({\bf{x}},{\bf{y}}) = \frac{1}{ \sqrt{V} } \sum_{ {\bf{k}} }
e^{ i ( {\bf{k}}-k_{F}{\hat{k}}) \cdot {\bf{x}} } e^{ i {\hat{k}}
\cdot {\bf{y}} }\mbox{ }c_{ {\bf{k}} } \mbox{
 }\theta(\Lambda - | |{\bf{k}}| - k_{F} | )
\label{PSIXY}
\end{equation}
First, this object is a slowly varying function of $ {\bf{x}} $.
In one dimension, a drastic simplification is possible which is
not available in more than one dimension, namely, $ e^{ i  sgn(k)
y } = \theta(k)  e^{ i y } + \theta(-k)  e^{ -i  y } $. In more
than one dimension, all possible directions $ {\hat{k}} $ are
involved. It is this $ \psi({\bf{x}}, {\bf{y}}) $ which is a
slowly varying function of $ {\bf{x}} $ that is easily expressed
in terms of the currents and densities rather than the full $
\psi({\bf{x}}) $.
 This field $ \psi({\bf{x}},{\bf{y}}) $ is the higher dimensional
 analog of the one dimensional version where only two directions are
 involved and would correspond to linear combinations of right and
 left movers.
Since this is a slowly varying function of $ {\bf{x}} $ we can
expect this to involve only $ \rho_{ {\bf{q}} } $ and $ X_{
{\bf{q}} } $ for small $ {\bf{q}} $ as is the case in one
dimension. Define $ \rho_{ {\bf{q}} } $ and $ X_{ {\bf{q}} }$ as
in Ref.\cite{Setlur1}. Then we have,
\begin{equation}
[\psi({\bf{x}}, {\bf{y}}), \rho_{ {\bf{q}} }]
 = \frac{1}{ \sqrt{V} } \sum_{ {\bf{k}} }
e^{ i ( {\bf{k}}-k_{F}{\hat{k}}) \cdot {\bf{x}} } e^{ i {\hat{k}}
\cdot {\bf{y}} }\mbox{ }c_{ {\bf{k-q}} } \mbox{
 }\theta(\Lambda - | |{\bf{k}}| - k_{F} | )
 \approx e^{ -i ( \nabla_{ {\bf{y}} }  \cdot {\bf{q}}) (\nabla_{ {\bf{y}} }  \cdot
{\bf{x}} ) } \mbox{     }\psi({\bf{x}}, {\bf{y}})
\end{equation}

\begin{equation}
[\psi({\bf{x}}, {\bf{y}}), X_{ {\bf{q}} }]
 =  \frac{1}{ \sqrt{V} } \sum_{ {\bf{k}} }
e^{ i ( {\bf{k}}-k_{F}{\hat{k}}) \cdot {\bf{x}} } e^{ i {\hat{k}}
\cdot {\bf{y}} }\mbox{ }
 \left( \frac{ i {\bf{p.q}} }{
N^{0} {\bf{q}}^2 } \right) \mbox{     }\delta_{ {\bf{k}}, {\bf{p}}
- {\bf{q}}/2 } c_{ {\bf{p}} + {\bf{q}}/2 }
 \mbox{
 }\theta(\Lambda - | |{\bf{k}}| - k_{F} | )
\approx e^{ i ( \nabla_{ {\bf{y}} }  \cdot {\bf{q}}) (\nabla_{
{\bf{y}} }  \cdot {\bf{x}} ) } \frac{ k_{F} {\bf{q}} \cdot
\nabla_{ {\bf{y}} } }{ N^{0} {\bf{q}}^2 } \mbox{
}\psi({\bf{x}}, {\bf{y}})
\end{equation}
These two rules may be exponentiated as follows.
\begin{equation}
\psi({\bf{x}}, {\bf{y}}) = e^{ -i \sum_{ {\bf{q}} } e^{ -i (
\nabla_{ {\bf{y}} }  \cdot {\bf{q}}) (\nabla_{ {\bf{y}} }  \cdot
{\bf{x}} ) } X_{ {\bf{q}} } } \mbox{          } e^{ i\sum_{
{\bf{q}} } e^{ i ( \nabla_{ {\bf{y}} }  \cdot {\bf{q}}) (\nabla_{
{\bf{y}} } \cdot {\bf{x}} ) }\left( \frac{ k_{F} {\bf{q}} \cdot
\nabla_{ {\bf{y}} } }{ N^{0} {\bf{q}}^2 } \right)\rho_{ {\bf{q}} }
} \mbox{ } F({\bf{x}}, {\bf{y}}) \label{PRES1}
\end{equation}
Here  $ F({\bf{x}}, {\bf{y}}) $ is an integration constant
independent of both $ X_{ {\bf{q}} } $ and $ \rho_{ {\bf{q}} } $.
It obeys the constraint that $ \nabla^2_{ {\bf{y}} } F({\bf{x}},
{\bf{y}}) = -F({\bf{x}}, {\bf{y}}) $
 since the unit vector $ {\hat{k}} $ multiplies $ {\bf{y}} $.
 In practical computations, one computes the propagator of the
interacting theory using the prescription in Eq.(\ref{PRES1}) and
multiplies and divides by the propagator of the free theory and in
the division uses the one obtained from Eq.(\ref{PRES1}) and in
the numerator uses the one obtained from elementary
considerations. This eliminates $ F({\bf{x}}, {\bf{y}}) $ and a
closed expression for the full dynamical (albeit slow part) of the
Green function may be written down in terms of current-current,
current-density, and density-density correlations.
 The full propagator may be related to
the current and density correlations as follows.
\[
\left<\psi^{\dagger}({\bf{x}}^{'},
{\bf{y}}^{'},t^{'})\psi({\bf{x}}, {\bf{y}},t)\right> = e^{
\frac{1}{2}\sum_{ {\bf{q}} } \left( \frac{ k_{F} {\bf{q}} \cdot
\nabla_{ {\bf{y}}^{'}} }{N^{0} {\bf{q}}^2 } \right)^2 \ll \rho_{
-{\bf{q}} }(t^{'}) \rho_{ {\bf{q}} }(t^{'})\gg } e^{
\frac{1}{2}\sum_{ {\bf{q}} } \left( \frac{ k_{F} {\bf{q}} \cdot
\nabla_{ {\bf{y}} } }{N^{0} {\bf{q}}^2 } \right)^2 \ll \rho_{
-{\bf{q}} }(t) \rho_{ {\bf{q}} }(t)\gg }
\]
\[
 e^{ -\sum_{ {\bf{q}} } \ll X_{ -{\bf{q}} }(t^{'}) X_{ {\bf{q}}
}(t^{'})\gg }
  e^{ -\sum_{ {\bf{q}}
}e^{ i ( \nabla_{ {\bf{y}} }  \cdot {\bf{q}}) (\nabla_{ {\bf{y}} }
\cdot {\bf{x}} ) } e^{ -i ( \nabla_{ {\bf{y}}^{'} }  \cdot
{\bf{q}}) (\nabla_{ {\bf{y}}^{'} }  \cdot {\bf{x}}^{'} ) }
 \left( \frac{ k_{F} {\bf{q}} \cdot \nabla_{
{\bf{y}}^{'} } }{ N^{0} {\bf{q}}^2 } \right)\ll \rho_{ -{\bf{q}}
}(t^{'}) X_{ -{\bf{q}} }(t)\gg }
\]
\[
e^{ \sum_{ {\bf{q}} } e^{ -i ( \nabla_{ {\bf{y}} }  \cdot
{\bf{q}}) (\nabla_{ {\bf{y}} } \cdot {\bf{x}} ) } e^{ i ( \nabla_{
{\bf{y}}^{'} }  \cdot {\bf{q}}) (\nabla_{ {\bf{y}}^{'} }  \cdot
{\bf{x}}^{'} ) }\left( \frac{ k_{F} {\bf{q}} \cdot \nabla_{
{\bf{y}} } }{ N^{0} {\bf{q}}^2 } \right) \ll X_{ -{\bf{q}}
}(t^{'}) \rho_{ -{\bf{q}} }(t)  \gg } e^{ \sum_{ {\bf{q}} } e^{ -i
( \nabla_{ {\bf{y}}^{'} }  \cdot {\bf{q}}) (\nabla_{ {\bf{y}}^{'}
}  \cdot {\bf{x}}^{'} ) }
 e^{ i ( \nabla_{ {\bf{y}} }  \cdot {\bf{q}}) (\nabla_{ {\bf{y}} }  \cdot
{\bf{x}} ) }
 \left( \frac{ k_{F}
{\bf{q}} \cdot \nabla_{ {\bf{y}} } }{ N^{0} {\bf{q}}^2 } \right)
\left( \frac{ k_{F} {\bf{q}} \cdot \nabla_{ {\bf{y}}^{'} } }{
N^{0} {\bf{q}}^2 } \right)
 \ll \rho_{ -{\bf{q}} }(t^{'})\rho_{ {\bf{q}} }(t) \gg }
 \]
\begin{equation}
  e^{ \sum_{ {\bf{q}} }e^{ i ( \nabla_{ {\bf{y}}^{'} }  \cdot
{\bf{q}}) (\nabla_{ {\bf{y}}^{'} }  \cdot {\bf{x}}^{'} ) }
 e^{ -i ( \nabla_{ {\bf{y}} }  \cdot {\bf{q}}) (\nabla_{ {\bf{y}} }  \cdot
{\bf{x}} ) } \ll X_{ -{\bf{q}} }(t^{'}) X_{ {\bf{q}} }(t) \gg }
\left<\psi^{\dagger}({\bf{x}}^{'},
{\bf{y}}^{'},t^{'})\psi({\bf{x}}, {\bf{y}},t)\right>_{0}
\label{PSIPSISLOW}
\end{equation}

where $ \ll ... \gg = < ... > - <...>_{0} $ and the subscript $
<...>_{0} $ refers to the quantities for the noninteracting
system. Next we wish to relate the  fundamental field $
\psi({\bf{x}}) $ to the slow field $ \psi({\bf{x}},{\bf{y}}) $.
This exercise is essential in order to ensure a logical
progression of ideas, also, it is the Green function of the
 fundamental field that is of physical significance.
 To accomplish this we may invert Eq.(\ref{PSIXY}) as follows.
\begin{equation}
\int \frac{ d {\bf{x}} }{ V } e^{ -i w{\hat{p}}.{\bf{x}} }
\int_{v_{0}} \frac{ d {\bf{y}} }{ v_{0} } e^{ -i
{\hat{p}}.{\bf{y}} }\psi({\bf{x}},{\bf{y}}) = \frac{1}{ \sqrt{V} }
\mbox{ }c_{ (k_{F}+w){\hat{p}} } \mbox{
 }\theta(\Lambda - |w|)
\end{equation}
Since the vector $ {\bf{y}} $ is dimensionless, we have to have an
arbitrary (formally infinite $ \sim (k_{F}L)^d $)
 dimensionless `volume' $ v_{0} $ to
 ensure that the transforms with respect to $ {\bf{y}} $ are properly inverted.
Thus the fundamental field is,
\begin{equation}
\psi({\bf{X}}) = \sum_{ {\bf{p}} }\int \frac{ d {\bf{x}} }{ V }
 e^{ i {\bf{p}}.{\bf{X}} }
 e^{ -i
(| {\bf{p}} |-k_{F}){\hat{p}}.{\bf{x}} } \int_{v_{0}} \frac{ d
{\bf{y}} }{ v_{0} } e^{ -i {\hat{p}}.{\bf{y}}
}\psi({\bf{x}},{\bf{y}}) = \frac{1}{ v_{0} } \int d {\bf{x}}
\int_{v_{0}} d{\bf{y}} \mbox{   }E({\bf{x}}, {\bf{y}}; {\bf{X}})
 \mbox{   }\psi({\bf{x}},{\bf{y}})
\label{PSIX}
\end{equation}
where $ E({\bf{x}}, {\bf{y}}; {\bf{X}}) = \frac{1}{V} \sum_{
{\bf{p}} } e^{ i {\bf{p}}.{\bf{X}} }
 e^{ -i
(p-k_{F}){\hat{p}}.{\bf{x}} } e^{ -i {\hat{p}}.{\bf{y}} } $. In
one and three spatial dimensions we have respectively,
\begin{equation}
E(x, y; X) \approx 2 \mbox{   }cos(k_{F}X-y)
 \mbox{   }\delta(X-x) \mbox{         };\mbox{      }
E({\bf{x}}, {\bf{y}}; {\bf{X}}) \approx \frac{ k_{F}^2 }{ 2\pi }
\mbox{ }\frac{ cos \left(\frac{
(k_{F}{\bf{X}}-{\bf{y}}).({\bf{X}}-{\bf{x}}) }{
|{\bf{X}}-{\bf{x}}| } \right)  }{ |{\bf{X}}-{\bf{x}}| }
\end{equation}
 In order for the one dimensional version of Eq.(\ref{PSIX})
 to be consistent with Eq.(\ref{EQPSIR}) and Eq.(\ref{EQPSIL}) we have to ensure that
 $ F(x,y) = \sqrt{ \frac{ 2\Lambda }{ \pi } } \mbox{           } cos(y) $.
 One may now write down the full propagator of the fundamental field in terms of
 current and density correlations by replacing the right hand side of the equation below
 by the expression for the slow propagator in Eq.(\ref{PSIPSISLOW}) as follows.
\begin{equation}
< \psi^{\dagger}({\bf{X}},t) \psi({\bf{X}}^{'},t^{'})>
 =  \frac{1}{ v_{0}^2 } \int d
{\bf{x}}^{'} \int_{v_{0}} d{\bf{y}}^{'} \mbox{   }E({\bf{x}}^{'},
{\bf{y}}^{'}; {\bf{X}}^{'})
 \mbox{   }\int d
{\bf{x}} \int_{v_{0}} d{\bf{y}} \mbox{   }E({\bf{x}}, {\bf{y}};
{\bf{X}})
 \mbox{   }
<\psi^{\dagger}({\bf{x}},{\bf{y}},t)
\psi({\bf{x}}^{'},{\bf{y}}^{'},t^{'})>
\end{equation}
We shall not dwell on this any further, for example, as a
nontrivial application, we could calculate the full dynamical
propagator of a two dimensional electron gas with long range
interactions (  $ V(r) = Log(r) $ ). This is bound to lead to the
conclusion that the system is a Luttinger liquid with a
characteristic exponent, a result that will be more easily and
elegantly shown using sea-bosons later on in this article. The
current and density correlations may be conveniently calculated
[at least at the RPA (random phase approximation)  level for
 translationally invariant systems] using the Lagrangian approach
that is conducive to the introduction of gauge
fields\cite{Pragauge}. To this end we derive an action functional
for the free electron system
 in terms of the hydrodynamic variables in any number of dimensions at the RPA level using
 the sea-boson formalism. Then we argue using general methods, that the functional form of the
 general action should be similar to the RPA result. First, the derivation of the RPA form
 of the action in terms of hydrodynamic variables. We borrow the definition of the hydrodynamic variables
 in terms of sea-bosons from our early work\cite{Setlur1}.
\begin{equation}
X_{ {\bf{q}} } = \frac{i}{ 2 N_{0} \epsilon_{ {\bf{q}} } }\sum_{
{\bf{k}} } \frac{ {\bf{k.q}} }{m} [ A_{ {\bf{k}} }({\bf{q}})
 + A^{\dagger}_{ {\bf{k}} }(-{\bf{q}}) ]
\mbox{          };\mbox{           } \rho_{ {\bf{q}} } = \sum_{
{\bf{k}} } [ A_{ {\bf{k}} }(-{\bf{q}})
 + A^{\dagger}_{ {\bf{k}} }({\bf{q}}) ]
\label{XRHO}
\end{equation}
We invert these using smearing functions. Let us postulate,
\begin{equation}
A_{ {\bf{k}} }({\bf{q}}) = -i \mbox{
}\Gamma_{X}({\bf{k}},{\bf{q}})
 X_{ {\bf{q}} }
 +  \Gamma_{\rho}({\bf{k}},{\bf{q}})
 \rho_{ -{\bf{q}} }
\label{smear}
\end{equation}
The above is meant to be valid in an average sense and is not an
operator identity. If we set,
\begin{equation}
\Gamma_{X}({\bf{k}},{\bf{q}}) = [A_{ {\bf{k}}
}({\bf{q}}),A^{\dagger}_{ {\bf{k}} }({\bf{q}})] \mbox{
};\mbox{          } \Gamma_{\rho}({\bf{k}},{\bf{q}})
 = \frac{1}{ 2N_{0} \epsilon_{ {\bf{q}} } } \frac{ {\bf{k.q}} }{m}
[A_{ {\bf{k}} }({\bf{q}}),A^{\dagger}_{ {\bf{k}} }({\bf{q}})]
\end{equation}
then substituting Eq.(\ref{smear}) into Eq.(\ref{XRHO}) leads to
an identity. We know from the sea-boson theory that the
hamiltonian of the free Fermi theory is
 $ H = \sum_{ {\bf{k}} {\bf{q}} } \frac{ {\bf{k.q}} }{m} A^{\dagger}_{ {\bf{k}} }({\bf{q}}) A_{ {\bf{k}} }({\bf{q}}) $.
Thus the action in terms of sea-bosons is,
\begin{equation}
S = \int^{ -i \beta }_{ 0 } dt \sum_{ {\bf{k}} {\bf{q}} }
A^{\dagger}_{ {\bf{k}} }({\bf{q}}) \left( i \partial_{t} - \frac{
{\bf{k.q}} }{m} \right)A_{ {\bf{k}} }({\bf{q}}) \label{acsea}
\end{equation}
Using Eq.(\ref{smear}) in Eq.(\ref{acsea}) we find,
\begin{equation}
S = \int^{ -i \beta }_{ 0 } dt \sum_{ {\bf{k}} {\bf{q}} }
 [2 \Gamma_{\rho}({\bf{k}},{\bf{q}}) \Gamma_{X}({\bf{k}},{\bf{q}})]\mbox{   }
 \rho_{ {\bf{q}} } \partial_{t}X_{ {\bf{q}} }
- \int^{ -i \beta }_{ 0 } dt \sum_{ {\bf{k}} {\bf{q}} }
 \Gamma_{\rho}({\bf{k}},{\bf{q}})^2
 \left( \frac{ {\bf{k.q}} }{m} \right)
 \rho_{ {\bf{q}} } \rho_{ -{\bf{q}} }
-\int^{ -i \beta }_{ 0 } dt \sum_{ {\bf{k}} {\bf{q}} }
\Gamma_{X}({\bf{k}},{\bf{q}})^2
 \left( \frac{ {\bf{k.q}} }{m} \right)
 X_{ {\bf{q}} }X_{ -{\bf{q}} }
\end{equation}
This may be simplified to yield,
\begin{equation}
 S = \int^{ -i \beta }_{ 0 } dt \sum_{ {\bf{q}} }
  \rho_{ {\bf{q}} } \partial_{t}X_{ {\bf{q}} }
- \int^{ -i \beta }_{ 0 } dt \sum_{ {\bf{q}} } \frac{ \epsilon_{0}
}{N_{0}}\mbox{   } \rho_{ {\bf{q}} } \rho_{ -{\bf{q}} } -\int^{ -i
\beta }_{ 0 } dt \sum_{ {\bf{q}} } N_{0} \epsilon_{ {\bf{q}}
}\mbox{   }
 X_{ {\bf{q}} }X_{ -{\bf{q}} }
\label{RPAACTION}
\end{equation}
where $ \epsilon_{0} = \sum_{ {\bf{k}} }
 \frac{3}{ 4N_{0} \epsilon_{ {\bf{q}} } }
n_{F}({\bf{k}})
 \left( \frac{ {\bf{k.q}} }{m} \right)^2 $. In one dimension
 $ \epsilon_{0} = \frac{ k^2_{F} }{ 2m } $.
 It appears
 that we have to use the one dimensional version of the energy scale $ \epsilon_{0} $ in higher dimensions as
 well to reproduce the right static structure factor. The reason for this is similar to the argument
 that allows Schotte and Schotte \cite{Schotte} to use the Tomonga-Luttinger theory to analyze X-ray absorption in bulk
 metals in three dimensions (only the s-wave contributes).
 We now argue that this action is the RPA limit of a more general formula for the action in terms
 of hydrodynamic variables that will be postulated rather than derived.
 The claim is that the general action  for the free
Fermi theory is given in terms of the current and density
variables as follows.
\begin{equation}
S = \int \left( \rho \partial_{t} \Pi -  V_{F}([\rho];{\bf{x}}) -
\frac{ \rho (\nabla \Pi)^2 + \frac{ ( \nabla \rho )^2 }{ 4 \rho }
}{ 2m } \right) \label{ACTION}
\end{equation}
Here $ \rho $ and $ \Pi $ are conjugate variables and $
{\bf{J}}({\bf{x}},t) = -\rho({\bf{x}},t) \nabla \Pi({\bf{x}},t) $
is the current. Also $ \rho({\bf{x}},t) = \frac{1}{V} \sum_{
{\bf{q}} n } e^{ -i {\bf{q.x}} } e^{ - z_{n} t } \rho_{ {\bf{q}} n
} $ and $ \Pi({\bf{x}},t) =\sum_{ {\bf{q}} n } e^{ i {\bf{q.x}} }
e^{ z_{n} t } X_{ {\bf{q}} n } $ where $ z_{n} = 2 \pi n/ \beta $
is the bosonic Matsubara frequency and $ \int \equiv \int^{ -i
\beta }_{0} dt \int d^d x $. Here $ V_{F}([\rho];{\bf{x}}) $ is a
suitable functional of the density chosen so as to reproduce the
properties of the free Fermi theory.
 To motivate the introduction of this new formalism we point out that
 in an earlier work\cite{Setlur1} one of the authors had argued that the fermionic
 field (operator) may be expressed in terms of the conjugate variable as
 follows.
\begin{equation}
\psi({\bf{x}},t) = e^{ i \Lambda([\rho];{\bf{x}},t) } e^{ -i
\Pi({\bf{x}},t) } \sqrt{ \rho({\bf{x}},t) } \label{PSIGRASS}
\end{equation}
 This means that the current operator may be written as follows $
 {\bf{J}}({\bf{x}},t) = - \rho({\bf{x}},t) \nabla \Pi({\bf{x}},t) + {\bf{C}}([\rho];{\bf{x}},t)
 $,  where $ {\bf{C}} $ depends on the $ \Lambda $ above. Such a
 general correspondence has already been suggested in the pioneering
 investigations of Rajagopal and Grest\cite{Raja}. An examination of
 the commutator $ [{\bf{J}}_{i}({\bf{x}},t),
 {\bf{J}}_{j}({\bf{x}}^{'},t)] $ in terms of the Fermi fields shows that it is expressible
 as linear combinations of terms involving the currents themselves.
 By imposing this requirement on the bosonic representation for the
 current, we find that $ {\bf{C}} \equiv 0 $. This in
 turn means that $ \Lambda $ should be independent of $ {\bf{x}} $ in order for
  current algebra to be respected. Unfortunately this conflicts with
  the requirement that $ \Lambda $ obey a certain recursion
  relation that was introduced in an earlier work \cite{Setlur1}.
 Thus we take the point of view that $ \Lambda $ depends weakly on $
 {\bf{x}} $ so that the current is still given by $ {\bf{J}} = -
 \rho \nabla \Pi $. However, the crucial $ \Lambda $ makes its
 appearance in the action as has been depicted in Eq.(\ref{ACTION}).
  The viewpoint advocated in this article is that we shall not be
  too insistent on verifying fermion commutation rules, rather
  this issue may be cleverly circumvented by forcing agreement
  with the correlation functions of the free theory and also
  by ensuring that the commutation rules of Fermi bilinears which are
  simpler, are recovered properly. Since $ V_{F} $ is uniquely
  determined by this approach, we may work backwards and determine
  $ \Lambda $ and then see if fermion commutation rules are indeed
  being obeyed. This is likely to be a formidable task and is
  therefore omitted. There are some subtleties associated with making the transition from the Hamiltonian description
 to the Lagrangian description. The principal of the these is the question of normal ordering.
 We note that the action in Eq.(\ref{acsea}) is properly normal ordered.
 The hydrodynamic variables in Eq.(\ref{PSIGRASS}) have to be expanded in terms of
 the sea-bosons and terms have to be rearranged so that all the annihilation operators are
 to the extreme right. Then one de-quantizes  Eq.(\ref{PSIGRASS}) by demoting the various operators
 to c-number functions before using it to derive the action. The claim is that the resulting
 ordinary complex number valued function `simulates' a Grassmann-valued field. The anticommuting
 property of the Grassmann variable is captured at the level of the propagator by the
 KMS(Kubo-Martin-Schwinger)-like boundary condition obeyed by the global Klein factor
 introduced in an earlier work\cite{Citrin}.
 As it is clear that normal ordering merely redefines $ \Lambda $, the actual nature
 of the function $ \Lambda $ is determined by making contact with the free Fermi theory.
 To this end, we may expand the action in Eq.(\ref{ACTION}) in
powers of density fluctuations. Retaining only the harmonic terms
amounts to using the RPA. At this level we may write,
\begin{equation}
\Lambda([\rho];{\bf{x}},t) \approx \sum_{ {\bf{q}}  \neq 0, n \neq
0 } e^{ i {\bf{q.x}} } e^{ z_{n} t }\mbox{        } \Lambda_{
{\bf{q}} n } \mbox{ } \rho_{ -{\bf{q}}, -n } \label{LAMBDA}
\end{equation}
where $ \Lambda_{ {\bf{q}} n } = \frac{ \epsilon_{F} }{ z_{n}
N^{0} } $ and $ N^{0} $ is the total number of electrons and $
\epsilon_{F} $ is the Fermi energy.
 Therefore $ V_{F}([\rho];{\bf{x}}) = \rho \partial_{t} \Lambda $.
It can be shown that the action in Eq.(\ref{ACTION}) reproduces
the right current and density correlations of the free Fermi
theory at the RPA level provided we choose $ \Lambda $ to be given
by Eq.(\ref{LAMBDA}).
 In particular the choice in Eq.(\ref{LAMBDA}) substituted into Eq.(\ref{ACTION}) yields the RPA level
 action Eq.(\ref{RPAACTION}) derived earlier rigorously using sea-bosons.
 One may then contemplate adding
more terms to this action such as density-density interactions,
density-phonon interactions or current density-vector potential
interactions to the action and expect to obtain nontrivial results
for the one-particle properties.
 We may also make Eq.(\ref{ACTION}) plausible by pointing out that it implies a connection
 between the generating function of density correlations and current correlations.
 This is relegated to the appendix.

In fact, we may go further and derive a general expression for the
generating function of density-density correlations of interacting
Fermi systems in terms of the corresponding correlation functions
of the free Fermi theory. This exercise is straightforward to
perform mainly because the action functional in Eq.(\ref{ACTION})
has a simple dependence on the conjugate variable $ \Pi $ which
may be integrated out. To this end, we may define the generating
function for the density-density correlation functions as,
\begin{equation}
Z([U]) = \int D[\rho] \mbox{       }\int D[\Pi] \mbox{     } e^{ i
S }  \mbox{       }e^{ \int^{ -i \beta }_{0} \int d^dx  \mbox{
}U({\bf{x}},t) \rho({\bf{x}},t) } \label{ZU}
\end{equation}
Correlation functions involving only the density variable may be
obtained as appropriate function derivatives with respect to the
source $ U $, for example, $ <T \mbox{    }[\rho({\bf{x}},t)
\rho({\bf{x}}^{'},t^{'})] > = \delta^2 Z([U])/ \delta
U({\bf{x}},t) \delta U({\bf{x}}^{'},t^{'})|_{ U \equiv 0 } $. The
equation above ( Eq.(\ref{ZU}) ) may be formally inverted and an
expression for the action with the conjugate variable integrated
out, may be written down in terms of the generating functional,
\begin{equation}
\int D[\Pi] \mbox{ } e^{ i S }
 = \int D[U^{'}] \mbox{        }e^{ -\int^{ -i \beta }_{0} \int d^dx \mbox{
}U^{'}({\bf{x}},t) \rho({\bf{x}},t) } Z([U^{'}])
\end{equation}
One may now use the above expression to relate the generating
function of the free theory with that of the interacting theory.
Thus,
\[
Z([U]) = \int D[\rho] \mbox{   }\int D[\Pi] \mbox{ } e^{ i
S_{free} } e^{ - \frac{i}{2}\int^{ -i \beta }_{0}dt \int d^d x
\int d^d x^{'} \mbox{ }V({\bf{x}}-{\bf{x}}^{'}) \rho({\bf{x}},t)
\rho({\bf{x}}^{'},t) } \mbox{       }e^{ \int^{ -i \beta }_{0}
\int d^dx  \mbox{ }U({\bf{x}},t) \rho({\bf{x}},t) }
\]
\begin{equation}
= \int D[U^{'}] \mbox{        } \int D[\rho]\mbox{          } e^{
- \frac{i}{2}\int^{ -i \beta }_{0}dt \int d^d x \int d^d x^{'}
\mbox{ }V({\bf{x}}-{\bf{x}}^{'}) \rho({\bf{x}},t)
\rho({\bf{x}}^{'},t) } e^{ \int^{ -i \beta }_{0} \int d^dx \mbox{
}(U({\bf{x}},t)  - U^{'}({\bf{x}},t) ) \rho({\bf{x}},t) }
Z_{free}([U^{'}])
\end{equation}
We may now decompose the density variable into the various Fourier
modes and perform the integration over the density variable and
write,
\begin{equation}
Z([U]) = \int D[U^{'}] \mbox{        } e^{ \sum_{ {\bf{q}} \neq 0
n } \frac{ V }{ 2 \beta V_{ {\bf{q}} } } (U_{ {\bf{q}} n } -
U^{'}_{ {\bf{q}},n } )(U_{ -{\bf{q}}, -n } - U^{'}_{ -{\bf{q}}, -n
} ) }\mbox{        } Z_{free}([U^{'}])
\end{equation}
where $ V_{ {\bf{q}} } = \int d^d x  \mbox{         }V({\bf{x}})
\mbox{         }e^{ -i {\bf{q}}.{\bf{x}} } $ and $ U_{ {\bf{q}} n
} =  \frac{1}{V}\int^{ -i \beta }_{0} dt \int d^dx \mbox{ }
U({\bf{x}},t)  e^{ -i {\bf{q.x}} } e^{ - z_{n} t } $. The
generating functional of the free theory may be written as
follows.
\begin{equation}
Z_{free}([U]) =
 e^{ \frac{1}{2!} \sum_{ {\bf{q}} \neq 0, n } \mbox{   } <\rho_{ {\bf{q}} n }\rho_{ -{\bf{q}},-n }
>_{0}\mbox{      } U_{ {\bf{q}} n } U_{ -{\bf{q}},-n } } \mbox{                } e^{ \frac{1}{3!}
\sum_{ {\bf{q}} n; {\bf{q}}^{'}, n^{'} } \mbox{   } <\rho_{
{\bf{q}} n }\rho_{ {\bf{q}}^{'}, n^{'} } \rho_{
-{\bf{q}}-{\bf{q}}^{'}, -n-n^{'} }
>_{0}\mbox{      } U_{ {\bf{q}} n } U_{ {\bf{q}}^{'}, n^{'} }
U_{ -{\bf{q}}-{\bf{q}}^{'}, -n-n^{'} } } ...
\end{equation}
Thus we may rightly suspect that retaining only the gaussian terms
in the above generating functional gives us the density-density
correlation in the sense of RPA. Corrections to RPA are obtained
by retaining the higher order terms involving three-body density
correlations. Unfortunately, the current-current correlations are
not so simple. This is because the current is a nonlinear
combination of the density and conjugate variables. However, one
may contemplate reverting to the Hamiltonian description and
invoking the equation of continuity to relate the current-current
correlations with the density-density correlations.

 Next we wish to show how to express the annihilation operator in
momentum space $ c_{{\bf{p}} } $ directly in terms of
sea-bosons\cite{Citrin} in contrast to the exercise just
completed, where we expressed only the slow part of the field in
terms of the current and densities. To be sure, the exercise in
this section is also valid only in the RPA-sense and hence only
provides the asymptotics. Define $ n_{ {\bf{q}} }({\bf{k}}) =
c^{\dagger}_{ {\bf{k}} - {\bf{q}}/2 } c_{ {\bf{k}} + {\bf{q}}/2 }
$ for $ {\bf{q}} \neq 0 $. Hence we may write,
\begin{equation}
[c_{ {\bf{p}} },n_{ {\bf{q}} }({\bf{k}})] = \delta_{ {\bf{p}},
{\bf{k}} - {\bf{q}}/2 } \mbox{ }c_{ {\bf{p+q}} } \equiv
 \delta_{ {\bf{p}}, {\bf{k}} - {\bf{q}}/2 }\mbox{          }
T_{ {\bf{q}} }({\bf{p}}) \mbox{   }c_{ {\bf{p}} } \label{RULE}
\end{equation}
where $ T_{ {\bf{q}} }({\bf{p}}) \equiv exp \left( {\bf{q}} \cdot
\nabla_{ {\bf{p}} } \right) $ is the translation operator. Also,
\begin{equation}
[n_{ {\bf{q}} }({\bf{k}}), n_{ -{\bf{q}}^{'} }({\bf{k}}^{'}) ] =
[c^{\dagger}_{ {\bf{k}} - {\bf{q}}/2 } c_{ {\bf{k}} + {\bf{q}}/2
},c^{\dagger}_{ {\bf{k}}^{'} + {\bf{q}}^{'}/2 } c_{ {\bf{k}}^{'} -
{\bf{q}}^{'}/2 }]
 \approx \left( n_{F}( {\bf{k}} - {\bf{q}}/2 ) - n_{F}( {\bf{k}} + {\bf{q}}/2
 ) \right) \mbox{    }\delta_{ {\bf{k}}, {\bf{k}}^{'} } \delta_{
 {\bf{q}}, {\bf{q}}^{'} }
  = sgn( {\bf{k.q}} )\mbox{    }\delta_{ {\bf{k}}, {\bf{k}}^{'} } \delta_{
 {\bf{q}}, {\bf{q}}^{'} }
\end{equation}
The last two approximations are equivalent to the random phase
approximation. The commutation rule in Eq.(\ref{RULE}) may be
exponentiated as follows,
\begin{equation}
c_{ {\bf{p}} } = e^{ -\sum_{ {\bf{q}} }
sgn({\bf{p.q}}+{\bf{q}}^2/2) \mbox{ }
 n_{ -{\bf{q}} }({\bf{p}}+{\bf{q}}/2)\mbox{          }
T_{ {\bf{q}} }({\bf{p}}) } \mbox{   }f( {\bf{p}} )
\end{equation}
where $ f({\bf{p}}) $ is an integration constant independent of $
n_{ {\bf{q}} }({\bf{k}}) $, the exact nature of which will be
discussed subsequently. To be sure there are additional terms
caused by the fact that the translational operator also acts on
these commutators and so on. The reasons why these issues are not
important are relegated to future publications. Suffice it to say
that this is consistent with the RPA. Now we wish to evaluate $
f({\bf{p}}) $. To this end we make the following
surmise\cite{Citrin},
\begin{equation}
f( {\bf{p}} ) = e^{ -i P_{0}( {\bf{p}} ) } \sqrt{ n_{0}( {\bf{p}}
) }
\end{equation}
Here $ n_{0}({\bf{p}}) = c^{\dagger}_{ {\bf{p}} } c_{ {\bf{p}} } $
and $ P_{0}({\bf{p}}) $ is canonically conjugate to $
n_{0}({\bf{p}}) $, in other words, $ [P_{0}({\bf{p}}),
n_{0}({\bf{p}}^{'})] = i \delta_{ {\bf{p}}, {\bf{p}}^{'} } $. This
ensures that $ [c_{ {\bf{p}} }, n_{0}({\bf{p}}^{'})] = c_{
{\bf{p}} } \mbox{ }\delta_{ {\bf{p}}, {\bf{p}}^{'} }$. We have
also set $ [n_{ {\bf{q}} }({\bf{k}}), n_{0}({\bf{p}})] = 0 $.
 In practical computations, we are going to set $ n_{0}({\bf{p}}) \approx
n_{F}({\bf{p}}) = \theta(k_{F}-|{\bf{p}}|) $. This means we have
to ensure that $ P_{0}( {\bf{p}} ) $ is formally infinite in order
for it to be a conjugate to $ n_{0}({\bf{p}}) $. Thus we redefine,
\begin{equation}
f( {\bf{p}} ) = e^{ -i N^{0} \xi( {\bf{p}} ) } \mbox{
        }n_{F}({\bf{p}})
\end{equation}
Here $ N^{0} $ is a large quantity (total number of particles) and
$ \xi $ is an arbitrary c-number function. This means it obeys,
\begin{equation}
Lt_{ N^{0} \rightarrow \infty } \mbox{           }
 e^{ -i N^{0} \xi( {\bf{p}}
) } \mbox{           }e^{ i N^{0} \xi( {\bf{p}}^{'} ) } = \delta_{
{\bf{p}}, {\bf{p}}^{'} }
\end{equation}
Unfortunately this prescription leaves out an essential part of
the dynamics. For example we expect that for the free theory, $
c_{ {\bf{p}} }(t) = e^{ -i \epsilon_{ {\bf{p}} } t } c_{ {\bf{p}}
}(0)$. This is not obeyed unless we retain $ P_{0}({\bf{p}}) $ as
an operator conjugate to $ n_{0}({\bf{p}}) $ \footnote{ Since $
n_{ {\bf{q}} }({\bf{k}}) $ varies slowly in time we may ignore it
for this aspect of the discussion. }. Thus we will have to put the
rapidly varying exponential in by hand during practical
computations. More precisely, we are going to multiply and divide
by the free propagator and use the bosonized version in the
denominator and the one obtained from elementary considerations in
the numerator thereby rendering this issue moot.
In terms of bosonic operators defined in our earlier
work\cite{Citrin}, $ n_{ {\bf{q}} }({\bf{k}}) = A_{ {\bf{k}}
}({\bf{q}}) + A^{\dagger}_{ {\bf{k}} }(-{\bf{q}}) $. Therefore, $
sgn({\bf{p.q}} + {\bf{q}}^2/2) n_{ -{\bf{q}}
}({\bf{p}}+{\bf{q}}/2) = - A_{ {\bf{p}}+{\bf{q}}/2 }(-{\bf{q}}) +
A^{\dagger}_{ {\bf{p}}+{\bf{q}}/2 }({\bf{q}}) $. In other words
finally we obtain,
\begin{equation}
c_{ {\bf{p}} } = e^{ \sum_{ {\bf{q}} \neq 0 } \left(A_{
{\bf{p}}+{\bf{q}}/2 }(-{\bf{q}}) - A^{\dagger}_{
{\bf{p}}+{\bf{q}}/2 }({\bf{q}})  \right)
 \mbox{          }
T_{ {\bf{q}} }({\bf{p}}) } \mbox{   } e^{ -i N^{0} \xi( {\bf{p}} )
} \mbox{
        }n_{F}({\bf{p}})
        \label{EQCP}
\end{equation}
 By expanding the exponential,
  it can be easily verified that this formula is consistent with the
 definition $ n_{ {\bf{q}} }({\bf{k}}) = c^{\dagger}_{
 {\bf{k}}-{\bf{q}}/2 } c_{ {\bf{k}} + {\bf{q}}/2 } $.
Notice that there are no Klein factors in this formula either.
Practical applications of the above formula will be relegated to
future publications. One may contemplate using Eq.(\ref{EQCP})
 to compute the one-particle Green function of any Fermi system.
  The results are valid at the level of RPA. In fact in an earlier
  work\cite{PraLutt}, one of the authors has highlighted the
  novel non-Fermi liquid character of a system described by
  long-range interactions in one dimension. In the following
  section, we study the same system in
  two spatial dimensions rather than in one. Here too we find that
  the system is an anomalous metal ( or an insulator ) but of the usual power-law Luttinger kind.

\section{Metal-Insulator Quantum Phase Transition in Two Dimensions}

Now we wish to turn our attention to the properties of a system of
spinless electrons mutually interacting via a gauge potential $
v_{ {\bf{q}} } = 2 \pi e^2/q^2 $ in two spatial dimensions. In
real space this corresponds to a $ Log(r) $ potential. Theories of
novel non-Fermi liquid properties in higher dimensions are rare
 one such study for example is by Gori-Giorgi and
Ziesche\cite{Gori}. We consider the following hamiltonian
 in the sea-boson language \cite{PraLutt}.
\begin{equation}
H = \sum_{ {\bf{k}} {\bf{q}} } \left( \frac{ {\bf{k.q}} }{m}
\right) A^{\dagger}_{ {\bf{k}} }({\bf{q}}) A_{ {\bf{k}}
}({\bf{q}})
 + \sum_{ {\bf{q}} \neq 0 } \frac{ v_{ {\bf{q}} } }{ 2 V}
 \sum_{ {\bf{k}} {\bf{k}}^{'} }[A_{ {\bf{k}} }(-{\bf{q}}) +
 A^{\dagger}_{ {\bf{k}} }({\bf{q}})]
[A_{ {\bf{k}}^{'} }({\bf{q}}) +
 A^{\dagger}_{ {\bf{k}}^{'} }(-{\bf{q}})]
\end{equation}
This represents electrons mutually interacting via a two body
potential $ v_{ {\bf{q}} } $ such that
 only small momentum transfer terms are important (forward scattering only) and
 $ {\bf{k}} $ and $ {\bf{k}}^{'} $ are close to the Fermi surface. Such a constraint is realized by the choice
 $ v_{ {\bf{q}} } = 2 \pi e^2/q^2 $ since the potential is singular for small $ q $ and
 negligible for larger $ q $. This hamiltonian may be easily diagonalized via a Bogoliubov transformation.
 We may borrow the results of Ando et al\cite{Stern} for the dielectric function of a two dimensional electron gas to
 arrive at the following formula for the dispersion of collective modes :
 $ \omega_{c}(q) \approx \frac{ e k_{F} }{ \sqrt{2m} } +
\left( \frac{ 3 k_{F} }{ 4 \sqrt{2} e m^{ \frac{3}{2} } } \right)
q^2 $.
 The smallness of $ q $ is governed by a cutoff $ \Lambda $ which is the value of $ q $ for which the second term
 is of the same order as the first. This means $ \Lambda \approx \frac{2}{ \sqrt{3} } e  \sqrt{m} $.
 The momentum distribution of this system at zero temperature may be evaluated using the sea-boson formulas\cite{PraLutt}.
 \begin{equation}
<n_{ {\bf{k}} }> = \frac{1}{ 2 } [ 1 + e^{ - 2 S^{0}_{B}({\bf{k}})
} ]
 n_{F}({\bf{k}}) + \frac{1}{ 2 } [ 1 - e^{ - 2 S^{0}_{A}({\bf{k}}) } ]
 (1-n_{F}({\bf{k}}))
\end{equation}
 where $ S^{0}_{A}({\bf{k}}) = \sum_{ {\bf{q}} }<A^{\dagger}_{ {\bf{k}}-{\bf{q}}/2 }({\bf{q}})A_{ {\bf{k}} - {\bf{q}}/2 }({\bf{q}})> $
 and $ S^{0}_{B}({\bf{k}}) = \sum_{ {\bf{q}} }<A^{\dagger}_{ {\bf{k}}+{\bf{q}}/2 }({\bf{q}})A_{ {\bf{k}}+{\bf{q}}/2 }({\bf{q}})> $
 and $ n_{F}({\bf{k}}) = \theta(k_{F}-|{\bf{k}}|) $. In the $ q $ summation we retain only the most singular parts
 and integrate only upto the cutoff $ \Lambda $.
 We omit the rest of the details regarding computation of the boson occupations.
 When this is done we obtain the following formula for the momentum
 distribution.
\begin{equation}
<n_{ {\bf{k}} }> = \frac{1}{2} + \frac{1}{2}  \left( \frac{
|k_{F}-|{\bf{k}}|| }{ \Lambda^{'} } \right)^{ \gamma }
 \mbox{        }sgn(k_{F}-|{\bf{k}}|)
\end{equation}
where $ \gamma = \frac{ m }{ k_{F} } \frac{ e }{ \sqrt{2m} } =
\frac{ \omega_{0} }{2E_{F}} $
 and $   \Lambda^{ '} = e^{ -\frac{ 2 }{ \pi } } \mbox{    }\Lambda  $ and
 $ \omega_{0} =  \frac{ e k_{F} }{ \sqrt{2m} } $. Thus the system an
 insulator of the usual Luttinger type for $ \omega_{0} > 2 E_{F} $ and an anomalous metal
 with a residual Fermi surface for $ \omega_{0} < 2 E_{F} $.
 We may now compute the dynamical Green function using the methods outlined in
 the earlier sections, but this is not going to yield any new information, since the important attributes namely
 the estimate of the plasmon energy and the anomalous exponent have already been highlighted.
 It would also be interesting to see if the usual Coulomb
 interaction in three spatial dimensions also exhibits some kind of
 metal insulator transition at absolute zero temperature as a function of
 density. Unfortunately in order to study this reliably we have to go
 beyond the RPA and this calculation is beyond the scope of this
 preliminary work that focuses mainly on formalism.

\section{ Conclusions }

To conclude, we have shown that it is possible to bosonize
fermions in arbitrary dimensions and relate bosonization with
quantum hydrodynamics. We have been able to extract the dynamical
one-particle Green function in terms of the current and density
correlation functions valid at the level of RPA. We have expressed
the fermion annihilation operator in momentum space directly in
terms of bosonic operators called sea-bosons. We have also derived
a general formula for the generating function of density
correlations of systems interacting via two-body forces in terms
of the corresponding quantity for the free theory. Finally we have
evaluated the momentum distribution of a two-dimensional anomalous
metal ( or insulator ) with the electrons interacting via a long
range potential and extracted the anomalous exponent.

\section{ Appendix }

 The connection between the current-current correlations and density-density correlation implied
  by the equation of continuity is recovered by the action Eq.(\ref{ACTION}) postulated in the main text.
   This may be verified at the level of RPA in any number of dimensions as follows.
 Define,
\begin{equation}
Z_{d}[U] \equiv \int D[ \rho ] D[ \Pi ] \mbox{    }
 e^{ i \int \left( \rho \partial_{t} \Pi -  \rho \partial_{t} \Lambda
- \frac{ \rho (\nabla \Pi)^2 + \frac{ ( \nabla \rho )^2 }{ 4 \rho
} }{ 2m } \right) + \int U \rho }
\end{equation}

\begin{equation}
Z_{c}[V] \equiv \int D[ \rho ] D[ \Pi ] \mbox{    }
 e^{ i \int \left( \rho \partial_{t} \Pi -  \rho \partial_{t} \Lambda
- \frac{ \rho (\nabla \Pi)^2 + \frac{ ( \nabla \rho )^2 }{ 4 \rho
} }{ 2m } \right) - \int {\bf{J}}\cdot \nabla V  }
\end{equation}

Then $ < T[ \rho({\bf{x}},t) \rho({\bf{x}}^{'},t^{'}) ] > = \frac{
\delta^2 Z_{d}[U] }{ \delta U({\bf{x}},t)\mbox{   } \delta
U({\bf{x}}^{'},t^{'}) } $  and  $ < T[ (\nabla_{ {\bf{x}} } \cdot
{\bf{J}}({\bf{x}},t)) (\nabla_{ {\bf{x}}^{'} }\cdot
{\bf{J}}({\bf{x}}^{'},t^{'})) ]
> = \frac{ \delta^2 Z_{c}[V] }{ \delta V({\bf{x}},t)\mbox{   }
\delta V({\bf{x}}^{'},t^{'}) } $. We may relate $ Z_{c} $ and $
Z_{d} $ as follows.

\begin{equation}
Z_{c}[V] = Z_{d}[- \frac{ i m }{ 2 } \mbox{        }(\nabla V)^2
+ m \mbox{
   }\partial_{t} V]
   \label{ZCV}
\end{equation}
The validity of this may be verified at the level of RPA. At this
level we may write,

\begin{equation}
Z_{d}[U] = e^{ \frac{1}{2} \int <T[ \rho({\bf{x}},t)
\rho({\bf{x}}^{'},t^{'}) ] >\mbox{    }
 U({\bf{x}},t) U({\bf{x}}^{'},t^{'}) }
\end{equation}
This implies,
\[
< T[ \nabla_{ {\bf{x}} } \cdot {\bf{J}}({\bf{x}},t) \mbox{    }
\nabla_{ {\bf{x}}^{'} } \cdot {\bf{J}}({\bf{x}}^{'},t^{'})  ] >
 =\int
 <T[ \rho({\bf{y}},T)
\rho({\bf{y}}^{'},T^{'}) ] >\mbox{    }
 ( m \mbox{   }\delta({\bf{y}}-{\bf{x}}) \partial_{T}\delta(T-t) )
( m \mbox{   }\delta({\bf{y}}^{'}-{\bf{x}}^{'})
\partial_{T^{'}}\delta(T^{'}-t^{'}) )
\]
\begin{equation}
= m^2 \mbox{   }<T[ \partial_{t} \rho({\bf{x}},t)
\partial_{t^{'}}\rho({\bf{x}}^{'},t^{'}) ]>
\end{equation}
The validity of the above equation follows from the equation of
continuity.

\section{Director's Cut}

In this section, we provide a new proof of the irrotational nature
of the velocity operator for fermions. This makes the above paper
logically complete. Unfortunately this proof is not there in the
published version.

\vspace{0.1in}

{\bf{Lemma}}: If $ \nabla \rho \times {\bf{J}} - \rho \nabla
\times {\bf{J}} = 0 $ then there exists a $ \Pi $ such that
 $ {\bf{J}} = - \rho \nabla \Pi $.

\vspace{0.1in}

{\bf{Proof}}: First we note that $
[{\bf{J}}({\bf{x}}),\rho({\bf{x}})] = 0 $, hence we may write $
\frac{ {\bf{J}} }{ \rho } $ without ambiguity.

\[
\nabla \times \frac{ {\bf{J}} }{ \rho } = \frac{1}{\rho} \nabla
\times {\bf{J}} - \frac{1}{ \rho^2 } \nabla \rho \times {\bf{J}} =
\frac{1}{ \rho^2 } [ \rho \nabla \times {\bf{J}} - \nabla \rho
\times {\bf{J}} ] = 0
\]

Hence $ \frac{ {\bf{J}} }{ \rho }  = - \nabla \Pi $ for some $ \Pi
$. Hence $ {\bf{J}} = - \rho \nabla \Pi $. Now we prove that $
\nabla \rho \times {\bf{J}} - \rho \nabla \times {\bf{J}} = 0  $
for (free) fermions. We do this using the first quantized
definitions of current and density operator and using real space
fermion wavefunctions.

\[
\rho({\vec{r}}) = \sum_{j=1}^{N} \delta({\vec{r}}-{\vec{r}}_{j})
\]
\[
{\bf{J}}({\vec{r}}) = \frac{i}{2}\sum_{j=1}^{N}
[\nabla\delta({\vec{r}}-{\vec{r}}_{j})]
 -i\sum_{j=1}^{N}
\delta({\vec{r}}-{\vec{r}}_{j})\nabla_{j}
\]
\[
\nabla\rho({\vec{r}}) \times {\bf{J}}({\vec{r}})= \sum_{j=1}^{N}
[\nabla\delta({\vec{r}}-{\vec{r}}_{j})] \times
\frac{i}{2}\sum_{p=1}^{N} [\nabla\delta({\vec{r}}-{\vec{r}}_{p})]
-\sum_{j=1}^{N} [\nabla\delta({\vec{r}}-{\vec{r}}_{j})] \times
i\sum_{p=1}^{N} \delta({\vec{r}}-{\vec{r}}_{p})\nabla_{p}
\]
\[
\rho({\vec{r}})\nabla \times {\bf{J}}({\vec{r}}) =
 -i
\sum_{j=1}^{N} \delta({\vec{r}}-{\vec{r}}_{j})
 \sum_{p=1}^{N}
[\nabla\delta({\vec{r}}-{\vec{r}}_{p})] \times \nabla_{p}
\]

\[
{\bf{V}}({\vec{r}}) = \sum_{j,p=1}^{N}
[\nabla\delta({\vec{r}}-{\vec{r}}_{j})] \times \frac{i}{2}
[\nabla\delta({\vec{r}}-{\vec{r}}_{p})]
-i\sum_{j,p=1}^{N}\delta({\vec{r}}-{\vec{r}}_{p})
[\nabla\delta({\vec{r}}-{\vec{r}}_{j})] \times
 \nabla_{p} + i \sum_{j,p=1}^{N}
\delta({\vec{r}}-{\vec{r}}_{j})
[\nabla\delta({\vec{r}}-{\vec{r}}_{p})] \times \nabla_{p}
\]

\[
{\bf{V}}({\vec{r}}) =  \frac{i}{2} \sum_{j,p=1}^{N} \left(
\delta({\vec{r}}-{\vec{r}}_{j})
[\nabla\delta({\vec{r}}-{\vec{r}}_{p})] -
\delta({\vec{r}}-{\vec{r}}_{p})
[\nabla\delta({\vec{r}}-{\vec{r}}_{j})]  \right) \times
(\nabla_{p} - \nabla_{j})
\]

Let, $ C({\vec{r}}_{j}, {\vec{r}}_{p}) \equiv \left(
\delta({\vec{r}}-{\vec{r}}_{j})
[\nabla\delta({\vec{r}}-{\vec{r}}_{p})] -
\delta({\vec{r}}-{\vec{r}}_{p})
[\nabla\delta({\vec{r}}-{\vec{r}}_{j})]  \right) \times
(\nabla_{p} - \nabla_{j})$ and $ f({\vec{r}}_{j}, {\vec{r}}_{p}) $
be some function :

\[
\int d {\vec{r}}_{j} \int d {\vec{r}}_{p} \mbox{      }
C({\vec{r}}_{j}, {\vec{r}}_{p})f({\vec{r}}_{j}, {\vec{r}}_{p}) =
\int d {\vec{r}}_{j} \int d {\vec{r}}_{p} \mbox{      } \left(
\delta({\vec{r}}-{\vec{r}}_{j})
[\nabla\delta({\vec{r}}-{\vec{r}}_{p})] -
\delta({\vec{r}}-{\vec{r}}_{p})
[\nabla\delta({\vec{r}}-{\vec{r}}_{j})]  \right) \times
({\vec{f}}_{2}({\vec{r}}_{j}, {\vec{r}}_{p}) -
{\vec{f}}_{1}({\vec{r}}_{j}, {\vec{r}}_{p}))
\]
\[
= \left( \int d {\vec{r}}_{p} \mbox{    }
[\nabla\delta({\vec{r}}-{\vec{r}}_{p})]\times
{\vec{f}}_{2}({\vec{r}}, {\vec{r}}_{p}) -\int d {\vec{r}}_{j}
[\nabla\delta({\vec{r}}-{\vec{r}}_{j})]\times
{\vec{f}}_{2}({\vec{r}}_{j}, {\vec{r}})   \right)
\]
\[
-\left( \int d {\vec{r}}_{p} \mbox{      }
[\nabla\delta({\vec{r}}-{\vec{r}}_{p})]\times
{\vec{f}}_{1}({\vec{r}}, {\vec{r}}_{p}) - \int d {\vec{r}}_{j}
[\nabla\delta({\vec{r}}-{\vec{r}}_{j})]\times
{\vec{f}}_{1}({\vec{r}}_{j}, {\vec{r}})  \right)
\]
\[
= \sum_{I} {\hat{e}}_{I} \mbox{          }\epsilon_{IJK} \mbox{
}\left( {\vec{f}}_{2,J;2,K}({\vec{r}}, {\vec{r}}) -
{\vec{f}}_{1,J;2,K}({\vec{r}}, {\vec{r}}) \right) -\sum_{I}
{\hat{e}}_{I} \mbox{          }\epsilon_{IJK} \left(
{\vec{f}}_{2,J;1,K}({\vec{r}}, {\vec{r}}) -
{\vec{f}}_{1,J;1,K}({\vec{r}}, {\vec{r}}) \right)
\]
\[
f_{1,J;2,K}({\vec{r}}, {\vec{r}})  = \left[ \frac{ \partial }{
\partial x_{J} } \frac{ \partial }{ \partial y_{K}
}f({\bf{X}},{\bf{Y}}) \right]_{ {\bf{X}}  = {\bf{Y}} = {\vec{r}} }
 = s \left[ \frac{ \partial }{
\partial x_{J} } \frac{ \partial }{ \partial y_{K}
}f({\bf{Y}},{\bf{X}}) \right]_{ {\bf{X}}  = {\bf{Y}} = {\vec{r}} }
= s f_{2, J ; 1, K }({\vec{r}},{\vec{r}})
\]

\[
f_{2,J;2,K}({\vec{r}}, {\vec{r}})  = \left[ \frac{ \partial }{
\partial y_{J} } \frac{ \partial }{ \partial y_{K}
}f({\bf{X}},{\bf{Y}}) \right]_{ {\bf{X}}  = {\bf{Y}} = {\vec{r}} }
 = s \mbox{   }\left[ \frac{ \partial }{
\partial y_{J} } \frac{ \partial }{ \partial y_{K}
}f({\bf{Y}},{\bf{X}}) \right]_{ {\bf{X}}  = {\bf{Y}} = {\vec{r}} }
= s f_{1,J;1,K}({\vec{r}}, {\vec{r}})
\]

\[
... = \left( f_{2,J;2,K}({\vec{r}}, {\vec{r}}) -
f_{1,J;2,K}({\vec{r}}, {\vec{r}}) \right) -\left(
f_{2,J;1,K}({\vec{r}}, {\vec{r}}) - f_{1,J;1,K}({\vec{r}},
{\vec{r}}) \right)
\]
\[
=  (s+1) f_{1,J;1,K}({\vec{r}}, {\vec{r}}) - (s+1) f_{2, J ; 1, K
}({\vec{r}},{\vec{r}})
\]
For $ s = -1 $ then these terms are zero.

\begin{acknowledgments}
The author would like to thank S. Basu for taking active interest
in the progress of this work and for various useful remarks that
enhanced the content of the work. It is also a pleasure to
acknowledge discussions with S. Das and S. Rao while one of the
authors (G.S.S.) was at the Harish Chandra Research Institute. An
early draft of this article was completed at this Institute.
\end{acknowledgments}


\end{document}